\documentclass[twocolumn,prb]{revtex4}
\usepackage{amsmath,amsthm}
\usepackage{epsfig}
\usepackage{graphics}
\usepackage{graphicx}
\usepackage{dcolumn}
\usepackage{bm}

\begin{document}

\title{Electronic correlations in the iron pnictides}
\author{M. M. Qazilbash,$^{1,\ast}$, J. J. Hamlin,$^1$ R. E. Baumbach,$^1$ Lijun Zhang,$^2$ D. J. Singh,$^2$
M. B. Maple,$^1$ and D. N. Basov$^1$}
\affiliation{$^1$ Physics
Department, University of California-San Diego, La Jolla,
California 92093, USA.\\ $^2$ Materials Science and Technology
Division, Oak Ridge National Laboratory, Oak Ridge, Tennessee
37831, USA.}

\date{\today}

\maketitle

\section{Text}

\textbf{In correlated metals derived from Mott insulators, the
motion of an electron is impeded by Coulomb repulsion due to other
electrons. This phenomenon causes a substantial reduction in the
electron's kinetic energy leading to remarkable experimental
manifestations in optical spectroscopy.\cite{millisreview} The
high-$T_c$ superconducting cuprates are perhaps the most studied
examples of such correlated metals. The occurrence of high-$T_c$
superconductivity in the iron pnictides \cite{hosono1,hosono2,ren}
puts a spotlight on the relevance of correlation effects in these
materials.\cite{si-abrahams,haule,laad} Here we present an
infrared and optical study on single crystals of the iron pnictide
superconductor LaFePO. We find clear evidence of electronic
correlations in metallic LaFePO with the kinetic energy of the
electrons reduced to half of that predicted by band theory of
nearly free electrons. Hallmarks of strong electronic many-body
effects reported here are important because the iron pnictides
expose a new pathway towards a correlated electron state that does
not explicitly involve the Mott transition.}

The recent discovery of superconductivity in the iron pnictides
promises to be an important milestone in condensed-matter
physics.\cite{hosono1,hosono2} Here is a new class of materials
with a layered structure and relatively high superconducting
\textit{T}$_c$ values \cite{hosono2,ren} rivalling the doped
cuprates. Electronic conduction is believed to occur in the
iron-pnictogen layers, \cite{hosono3} similar to the cuprates
where the charge carriers are delocalized in the copper-oxygen
planes. Two decades of research on the cuprates has established
that a proper account of the exotic normal state properties is a
prerequisite for the understanding of the superconducting
instability.\cite{bonn} Thus motivated, we investigated the normal
state of the iron pnictide superconductors with infrared and
optical spectroscopy, focusing on charge dynamics in the
conducting planes.

\begin{figure}[t]
\epsfig{figure=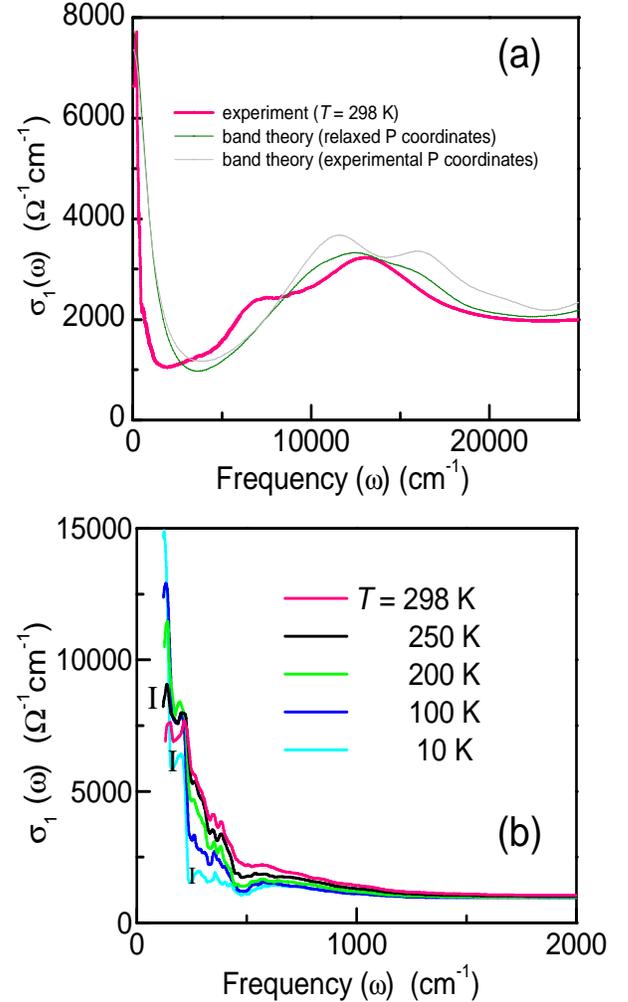,width=80mm,height=135mm}
\caption{(a) Real part of the \textit{ab}-plane optical
conductivity $\sigma_1$($\omega$) of LaFePO is plotted as a
function of frequency for \textit{T} = 298 K. Also shown are
$\sigma_1$($\omega$) plots calculated within band theory. (b)
Systematic temperature dependence of $\sigma_1$($\omega$). The
uncertainty in $\sigma_1$($\omega$) in panel (b) is less than or
equal to the thickness of the lines for $\omega$ $>$ 350
cm$^{-1}$. However, the uncertainty in $\sigma_1$($\omega$)
increases at lower frequencies as indicated by the error
bars.}\label{sigma1}
\end{figure}

An optical experiment measures the dynamical response of the
electron subjected to an external electromagnetic field and
facilitates monitoring of many-body effects experienced by the
electron in a material. These many-body effects include the
interaction of the electron with other electrons, phonons as well
as ordered or fluctuating spins. In Fig.~\ref{sigma1}a we display
the real part of the $ab$-plane optical conductivity
$\sigma_1$($\omega$) of LaFePO over a broad frequency range.
Sample growth and characterization procedures, and the
experimental details for obtaining the optical conductivity are
provided in Methods and Supporting Information. The low frequency
infrared response is dominated by the narrow Drude feature
signifying the presence of itinerant charge carriers. Two distinct
interband transitions in the form of hump-like structures are seen
at higher frequencies. The temperature dependence of the Drude
feature is displayed in Fig.~\ref{sigma1}b. The Drude peak at low
frequencies becomes sharper, and the conductivity at the lowest
measured frequencies increases with decreasing temperature which
is typical metallic behavior.

\begin{figure}[t]
\epsfig{figure=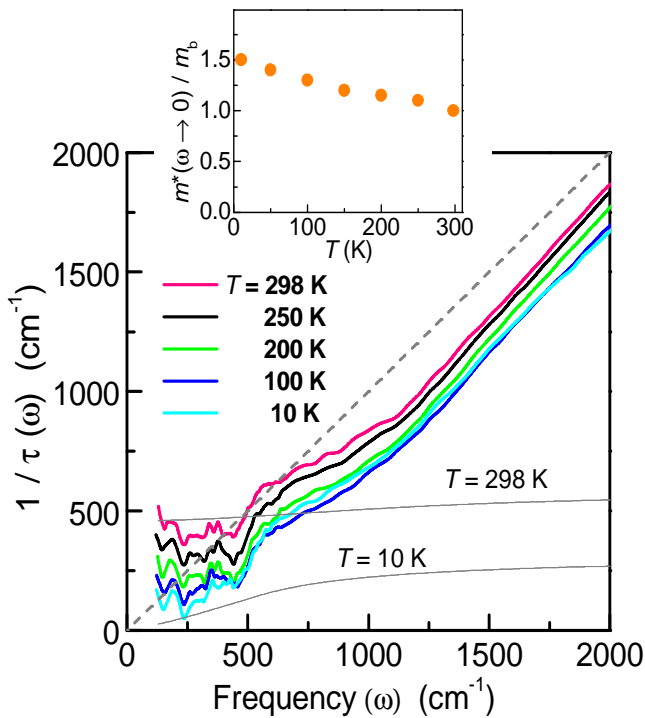,width=85mm,height=95mm}
\caption{The frequency dependence of the scattering rate
1/$\tau$($\omega$) is plotted for representative temperatures in
the normal state of LaFePO.  The dashed line represents the
equation $\omega$ = 1/$\tau$($\omega$). The thin solid lines are
the expected electron-phonon scattering rate at $T$ = 298 K and
$T$ = 10 K calculated using a model phonon spectral density for
LaFePO (see Supporting Information). The inset shows the
temperature dependence of the mass enhancement factor in the low
frequency limit \textit{m}*($\omega$ $\rightarrow$
0)/\textit{m}$_b$.} \label{xtendedDrude}
\end{figure}

In order to gain insight into the effect of many-body interactions
on the itinerant behavior of the charge carriers, we performed the
extended Drude analysis~\cite{basovreview}. Via the extended Drude
analysis, we extract the scattering rate 1/$\tau$($\omega$) and
the mass enhancement factor \textit{m}*($\omega$)/\textit{m}$_b$
(see Supporting Information). The temperature dependence of
1/$\tau$($\omega$) and $m^*$($\omega \rightarrow$ 0)/$m_b$ is
respectively plotted in Fig.~\ref{xtendedDrude} and the inset. One
can see that the scattering rate increases continuously and
monotonically with frequency. Electron-phonon scattering theory
predicts a constant, frequency-independent scattering rate above
the cutoff of the phonon spectrum \cite{basovreview,VO2}.
Measurements and calculations of the phonon spectral density for
LaFePO are not yet available. However, the cutoff is not likely to
exceed 550 cm$^{-1}$ inferred from calculations on the iron
oxy-arsenides \cite{djsingh,boeri}. Therefore, the absence of
saturation of 1/$\tau$($\omega$) at higher frequencies in LaFePO
suggests that in addition to electron-phonon scattering, there are
other scattering channel(s) arising from electronic and spin
correlations. It is interesting to note that the scattering rate
has a linear frequency dependence between 1100 cm$^{-1}$ and 2000
cm$^{-1}$ and it deviates from linearity below 1100 cm$^{-1}$.
This frequency dependence almost mimics the temperature dependence
of the resistivity which has a linear temperature dependence at
higher temperatures and a quadratic temperature dependence at low
temperatures.\cite{maple}

At $T$ = 10 K, the frequency dependent scattering rate lies below
the line $\omega$ = 1/$\tau$($\omega$) at low frequencies,
signifying the presence of coherent quasiparticles. This is
consistent with the observation of deHaas-vanAlphen oscillations
in LaFePO.\cite{dHvA} Note, however, that 1/$\tau$($\omega$) is of
the order of the excitation energy (frequency) at high energies
(frequencies) even at low temperatures. This is significant since
quasiparticles may be well-defined only near the Fermi energy
whereas at higher energies we are dealing with a strongly
dissipative system. At higher temperatures, 1/$\tau$($\omega$)
approaches the line $\omega$ = 1/$\tau$($\omega$) i.e. the
incoherent regime. In doped iron-arsenides, the scattering rate
exceeds the excitation frequency,\cite{timusk} implying that
transport in the iron-arsenides is more incoherent than in its
phosphorous based counterparts. We note that the mass enhancement
in the zero frequency limit due to low energy many-body
interactions is at most 1.5 (inset of Fig.~\ref{xtendedDrude}).
This value of the mass enhancement is obtained using the
experimentally determined plasma frequency which is already
renormalized by high energy electronic correlations (see
subsequent discussion and Supporting Information). Therefore, the
mass enhancement obtained from the extended Drude analysis is a
result of low energy many-body interactions. On the other hand, we
find that high energy electronic correlations lead to an effective
mass renormalization in LaFePO by a factor of 2 which is
equivalent to a kinetic energy reduction by a factor of 2 compared
to the band theory value which we discuss next.

With itinerant nature of LaFePO firmly identified by our data in
Figs.\ref{sigma1} and \ref{xtendedDrude}, we now proceed to the
central part of our analysis pertaining to the electronic kinetic
energy. The optical conductivity data provides a measure of the
kinetic energy $K$ of the electrons.\cite{millisreview} The
kinetic energy can also be readily evaluated using band structure
calculations. As a rule, experimental results for itinerant
electron systems are in good agreement with the band structure
calculations leading to $K_{exp}$/$K_{band}$ close to unity in
simple metals (see Fig.~3). However, interesting electronic
effects due to strong interactions involving charge, spin and
orbital degrees of freedom occurring in many intermetallic
compounds with $d$- (and $f$-) electrons yield dynamical effects
beyond ordinary band structure results. All these effects compete
with itinerancy of electrons leading to the suppression of
$K_{exp}$/$K_{band}$ value from unity. The most notorious example
is a Mott insulator where $K_{exp}$ = 0 due to Coulomb repulsion
whereas band structure results still predict metallic response
with finite $K_{band}$. A trend revealed by metals in the vicinity
of Mott insulators is that $K_{exp}$/$K_{band}$ increases from
zero to finite values as Coulomb correlations are suppressed
resulting in a Mott insulator-to-metal transition with doping or
temperature (Fig.3). While $K_{exp}$/$K_{band}$ in correlated
metals derived from Mott insulators is finite, it is significantly
less than unity. Therefore optical experiments in tandem with band
structure results offer reliable means to probe electronic
correlations in materials.

We have calculated $\sigma_1$($\omega$) from band theory within
the generalized gradient approximation (GGA) using both
experimental and relaxed coordinates (see Fig.~\ref{sigma1}a and
Supporting Information). The calculated $\sigma_1$($\omega$) shows
a Drude feature at low frequencies and higher frequency interband
transitions, in reasonable agreement with the experimental data.
The band theory calculations show that LaFePO is a metal in
agreement with experiment. There are, however, some differences
between the calculated and measured $\sigma_1$($\omega$) that will
be elaborated in subsequent paragraphs. We note here that GGA
calculations were performed with two choices of the atomic
coordinates. The two choices do not significantly affect the Drude
part but do affect the peak positions of the interband
transitions.

We now wish to draw attention to the key parameter associated with
the optical conductivity - the area under the Drude part of
$\sigma_1$($\omega$) which is proportional to the electron's
kinetic energy. One can immediately see in Fig.~\ref{sigma1}a that
the area under the Drude part of the measured $\sigma_1$($\omega$)
is significantly less than that calculated within band theory. To
put this on a quantitative footing, we define the experimental
kinetic energy as:\cite{millisndoped}

\begin{align}
K_{exp}(\omega_c)=\frac{{\hbar}c_0}{e^2}\int_{0}^{\omega_c}\frac{2\hbar}{\pi}{\sigma_1(\omega}){d\omega}
\label{energy}
\end{align}

In the above equation, $c_0$ is the distance between the FeP
planes. Upon integrating the Drude part of the experimental
$\sigma_1$($\omega$) up to a cutoff frequency $\omega_c$ = 3000
cm$^{-1}$, we obtain a kinetic energy $K_{exp}$ = 0.15 eV. The
theoretical value of the kinetic energy is obtained directly from
band theory calculations of the plasma frequency, which give
$K_{band}$ = 0.29 eV (see Supporting Information for details).
$K_{exp}$ is nearly 50\% of $K_{band}$ and this reduction is due
to correlation effects not accounted in band theory. We note that
this discrepancy between $K_{exp}$ and $K_{band}$ is robust with
respect to the uncertainties in the experimental and theoretical
values of the kinetic energy which are discussed in the Methods
section and Supporting Information. Furthermore, our findings are
in agreement with a recent photoemission experiment in which the
measured bandwidth is nearly half of that calculated within band
theory.\cite{zxshen}

Among the various superconducting iron-pnictide families, LaFePO
has one of the highest normal state conductivities. Despite the
high conductivity, the reduction of the kinetic energy is
significant. Therefore, the ground state wave-function of LaFePO
is not the same as calculated by band theory, and dynamical
correlation effects omitted by band theory have to be taken into
account for a realistic theoretical description. This view is
further supported by the differences between the measured optical
interband transitions and those derived from the band structure
(see Fig.~\ref{sigma1}a). In the experimental data, there are two
distinct peaks in $\sigma_1$($\omega$) centered at 7200 cm$^{-1}$
and 13000 cm$^{-1}$, most likely arising from optical transitions
between the Fe-3$d$ bands. We note that the band theory
calculations predict two peaks between 10000 and 17000 cm$^{-1}$
whose positions are sensitive to the atomic coordinates. The
discrepancy between the experimental data and band structure
calculations are likely due to modification of the actual band
structure arising from electronic correlations.

\begin{figure}[t]
\epsfig{figure=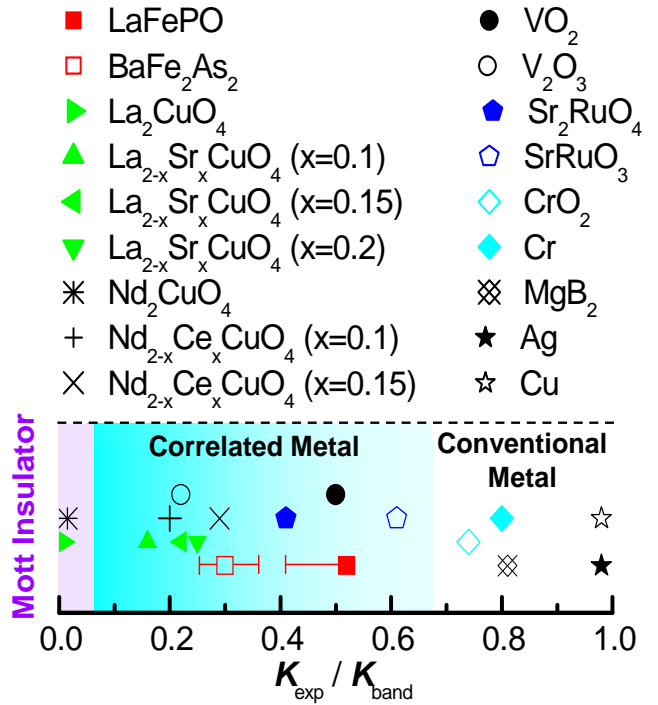,width=85mm,height=95mm}
\caption{The ratio of the experimental kinetic energy and the
kinetic energy from band theory $K_{exp}$/$K_{band}$ for the iron
pnictides and various other metals is plotted. The data points are
offset in the vertical direction for clarity. The value of the
LaFePO data point and the associated error bar are discussed in
the text, Methods section and the Supporting Information. The
values $K_{exp}$ and $K_{band}$ are obtained from the following
references: paramagnetic BaFe$_2$As$_2$
(Refs.\onlinecite{wang,ma,drechsler}), La$_{2-x}$Sr$_x$CuO$_4$
(Ref.\onlinecite{millisnatphys}), Nd$_{2-x}$Ce$_x$CuO$_4$
(Ref.\onlinecite{millisnatphys}), VO$_2$ rutile metal
(Ref.\onlinecite{VO2}), V$_2$O$_3$ paramagnetic metal
(Refs.\onlinecite{thomas,kotliar}), Sr$_2$RuO$_4$
(Refs.\onlinecite{tokura,djsingh2}), paramagnetic SrRuO$_3$
(Refs.\onlinecite{noh,mazin2}), half-metallic CrO$_2$
(Refs.\onlinecite{singley,pballen,mazincro2}), paramagnetic Cr
(Ref.\onlinecite{vdmarel}), MgB$_2$ ($a$-axis)
(Ref.\onlinecite{mgb2}), Ag and Cu (Ref.\onlinecite{agcu}). The
error bar on the BaFe$_2$As$_2$ data point is based on the scatter
in the theory and experimental values of the plasma frequency
(Ref.\onlinecite{wang,ma,drechsler}). The error bars (not shown)
for other materials are estimated to be at least the width of the
symbols and at most $\pm$ 0.05.} \label{diagram}
\end{figure}

The reduction of the kinetic energy compared to band theory in
LaFePO, its proximity to the incoherent metal regime, and the
absence of a Mott insulator in its phase diagram inspired us to
compare it to other exotic conductors that have been studied in
the past. Fig~\ref{diagram} is a map showing the location of
various exotic conductors as well as a few conventional metals (Ag
and Cu) based upon the ratio of the experimental kinetic energy to
that calculated within band theory. The exotic conductors that
have been selected for comparison share at least one of the
following features in common with the iron-pnictides: high-$T_c$
superconductivity, itinerant magnetism, and/or electronic
correlations. Within the ambit of Fig~\ref{diagram}, the
correlated metal regime is characterized by a substantially
reduced empirical kinetic energy of charge carriers compared to
theory. Note that electron-phonon coupling does not result in a
significant reduction of the electron's kinetic energy (unless it
is extremely strong as to lead to polaron
formation).\cite{millisreview} Electron-phonon coupling in MgB$_2$
superconductor, for example, results in $K_{exp}$/$K_{band}$ not
much less than unity in the metallic state (Fig.\ref{diagram}).

We now look at the consequences of magnetic interactions on the
kinetic energy. In paramagnetic Cr and half-metallic CrO$_2$, two
classic systems with itinerant magnetism, spin correlations lead
to 20-25\% reduction in $K_{exp}$ relative to $K_{band}$
(Fig.\ref{diagram}). Moreover, in the cuprates where the magnetism
has local moment character, opposite to the itinerant case
discussed above, magnetic interactions also lead to a relatively
small reduction in the electron's kinetic energy compared to that
predicted by band theory.\cite{millisnatphys} Even where it is
argued that the role of magnetic interactions in the cuprates is
important, it is seen that 80-90\% of the experimental reduction
in kinetic energy compared to band theory is accounted for by the
intra-atomic Coulomb repulsion and 10-20\% may be attributed to
antiferromagnetic correlations.\cite{millisnatphys} It is worth
emphasizing that $K_{exp}$ is reduced substantially compared to
$K_{band}$ in the cuprates, oxides of vanadium and the ruthenate
family of metals as shown in Fig.\ref{diagram} - these materials
have Mott insulators in their phase diagrams and dynamical Coulomb
correlations dominate transport behavior in the metallic states.
Thus, intra-atomic Coulomb repulsion is one established mechanism
that leads to a substantial reduction in $K_{exp}$ compared to
$K_{band}$.

We note that the parent compounds of the superconducting
iron-arsenides exhibit magnetic ordering at low temperatures,
\cite{dai} but they remain metallic and are not insulating in the
ordered state. Furthermore, LaFePO shows no signs of magnetic
ordering~\cite{uemura} or insulating behavior.\cite{maple}
However, spin correlations likely exist in LaFePO and the
paramagnetic phases of the iron arsenides.~\cite{djsingh} The data
in Fig~\ref{diagram} show that both in LaFePO and paramagnetic
BaFe$_2$As$_2$, $K_{exp}$/$K_{band}$ is less than that in other
typical itinerant magnets and is substantially less than unity.
Notice that $K_{exp}$/$K_{band}$ is further reduced for
paramagnetic BaFe$_2$As$_2$ compared to LaFePO which means that
the arsenides are even more correlated than the
phosphides.\cite{si-abrahams} These observations may not be
explained by spin correlations alone, in which case they signify
the relevance of dynamical Coulomb correlations to the physics of
the iron pnictides. This is remarkable in the absence of a Mott
transition in their phase diagrams. Therefore, the iron pnictides
are properly classified as being in the moderate correlation
regime with the on-site Coulomb repulsion of the order of the
bandwidth. There is the possibility that complexity related to
charge, magnetic and orbital degrees of freedom leads to a
correlated metal that need not be derived from a Mott insulator.
Our results demonstrate that transport in the iron-pnictides lies
between the band-like itinerant and the Mott-like local magnetic
moment extremes.

\section{Methods}

The growth and characterization procedures for LaFePO single
crystals are given in Ref.~\onlinecite{maple}. Resistivity and
magnetization measurements reveal a superconducting \textit{T}$_c$
of $\approx$ 6 K, with complete zero-field-cooled diamagnetic
shielding in the superconducting state. The crystals are
platelets, typically 0.5 mm $\times$ 0.5 mm $\times$ 0.05 mm in
size.

The \textit{ab}-plane reflectance was measured in the near-normal
incidence geometry in a Bruker v66 Infrared Fourier-transform
Spectrometer at frequencies between 100 cm$^{-1}$ and 24000
cm$^{-1}$. The reflectance measurements were performed at the
following temperatures: 298 K, 250 K, 200 K, 150 K, 100 K, 50 K,
and 10 K. We obtained the optical constants through fitting the
reflectance data with Drude and Lorentzian oscillators, and
Kramers-Kronig constrained variational dielectric functions as
described in Ref.~\onlinecite{kkvariational}. In addition,
variable-angle spectroscopic ellipsometry was performed in the
frequency range 5500 - 25000 cm$^{-1}$ which improves the accuracy
of the extracted optical constants in this frequency range. The
temperature dependence of the \textit{ab}-plane reflectance of
LaFePO crystals is displayed in Fig.S1 in the Supporting
Information.

Details regarding the extended Drude analysis are given in
Supporting Information online. We note here that
$K_{exp}$/$K_{band}$ = 0.52 for LaFePO plotted in
Fig.~\ref{diagram} is an upper bound, and other methods for
obtaining $K_{exp}$/$K_{band}$ give a lower bound of 0.4. Further
details about calculations of $K_{exp}$ and $K_{band}$ for LaFePO
are given in the Supporting Information. Also included in the
Supporting Information are calculations of the scattering rates
due to electron-phonon coupling and some notes on
Fig.~\ref{diagram}.

\section{Acknowledgements}

The authors are grateful to Elihu Abrahams, A. V. Boris, A. V.
Chubukov, A. J. Millis, Oleg Shpyrko, Qimiao Si, and Congjun Wu
for discussions. M.M.Q. thanks A. Kuzmenko for assistance with the
software for infrared data analysis. This work was supported in
part by the National Science Foundation grant NSF DMR 0705171.

\newpage

\begin{figure}[t]
\epsfig{figure=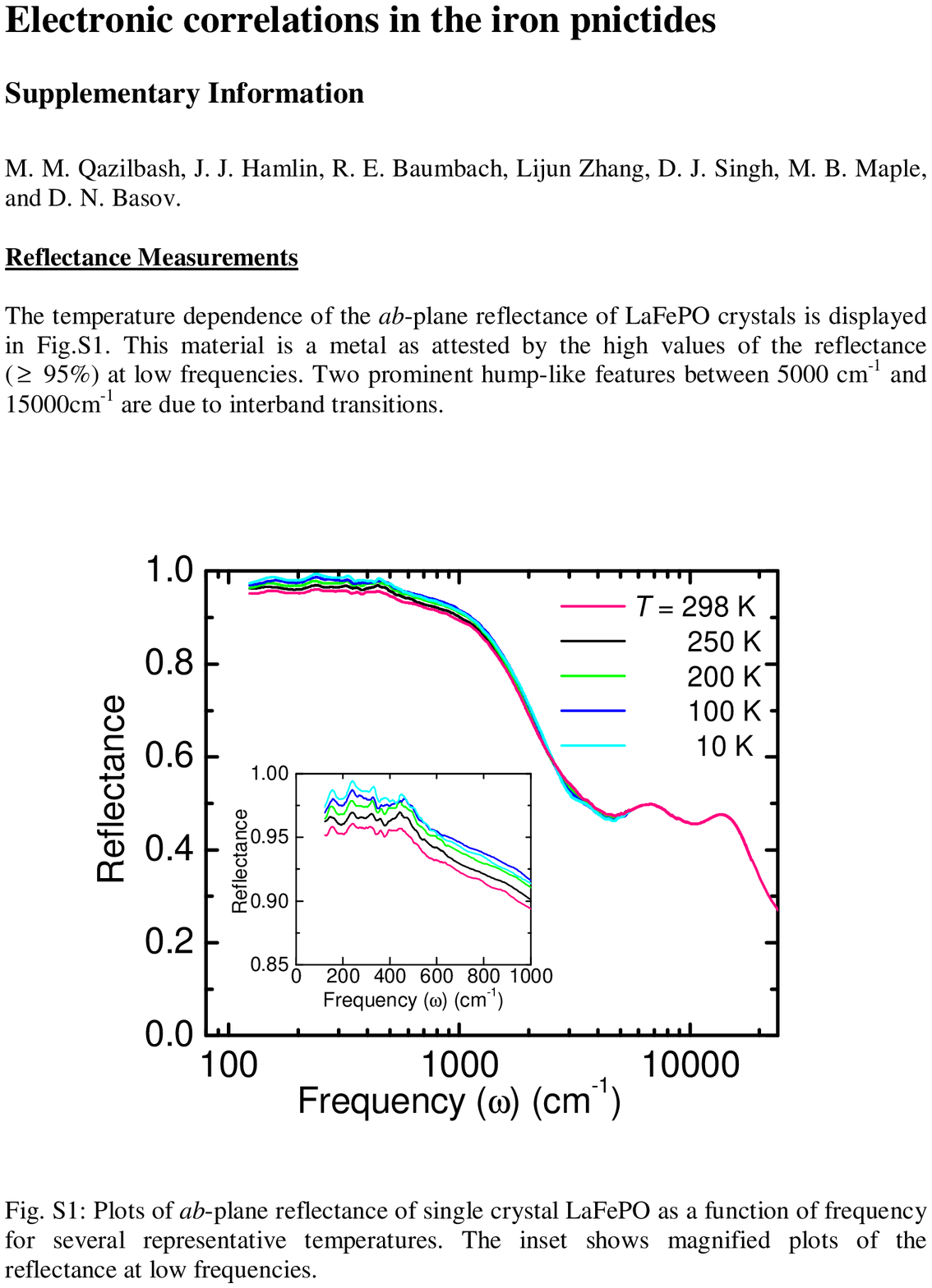}
\end{figure}

\newpage

\begin{figure}[t]
\epsfig{figure=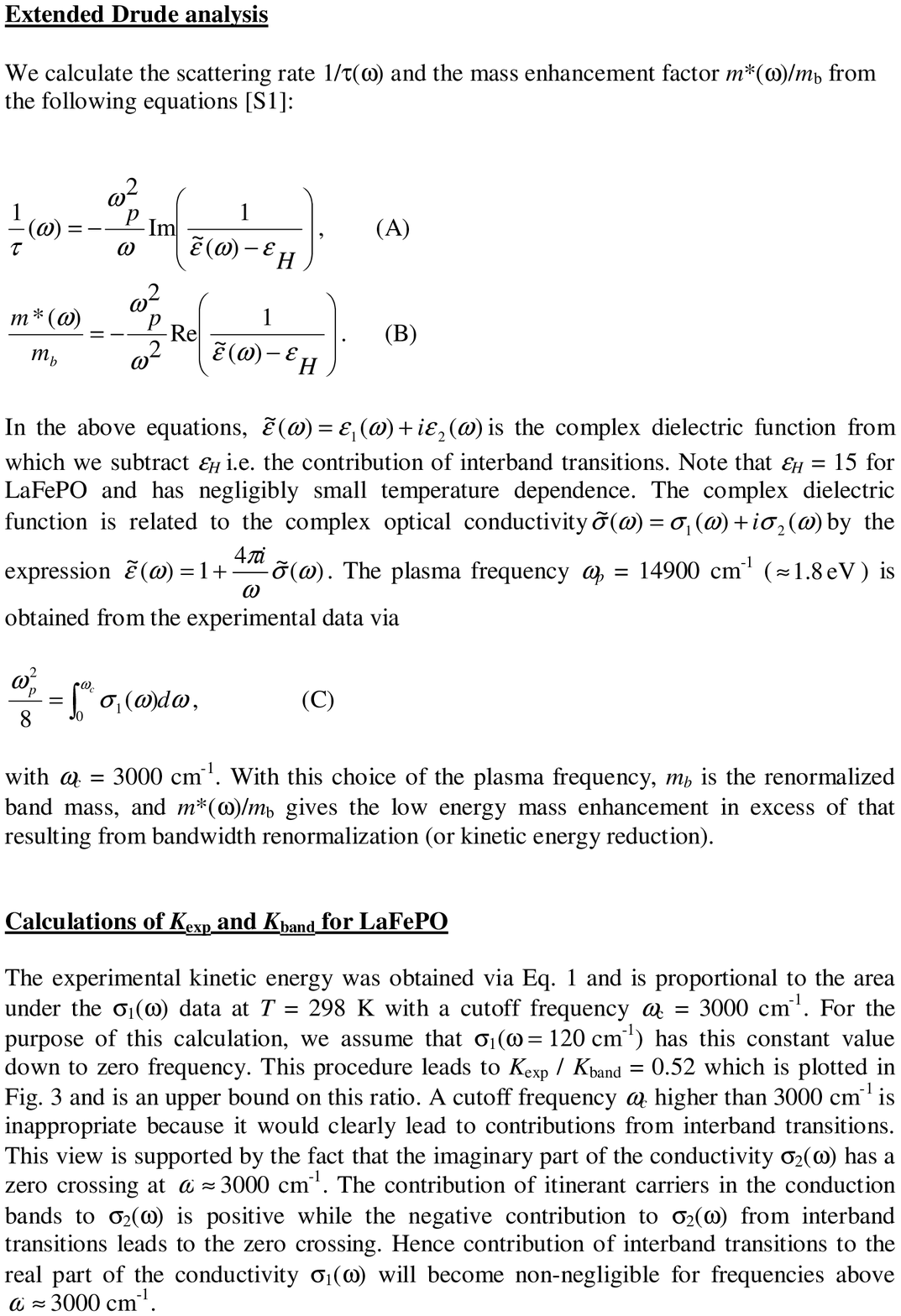}
\end{figure}

\newpage

\begin{figure}[t]
\epsfig{figure=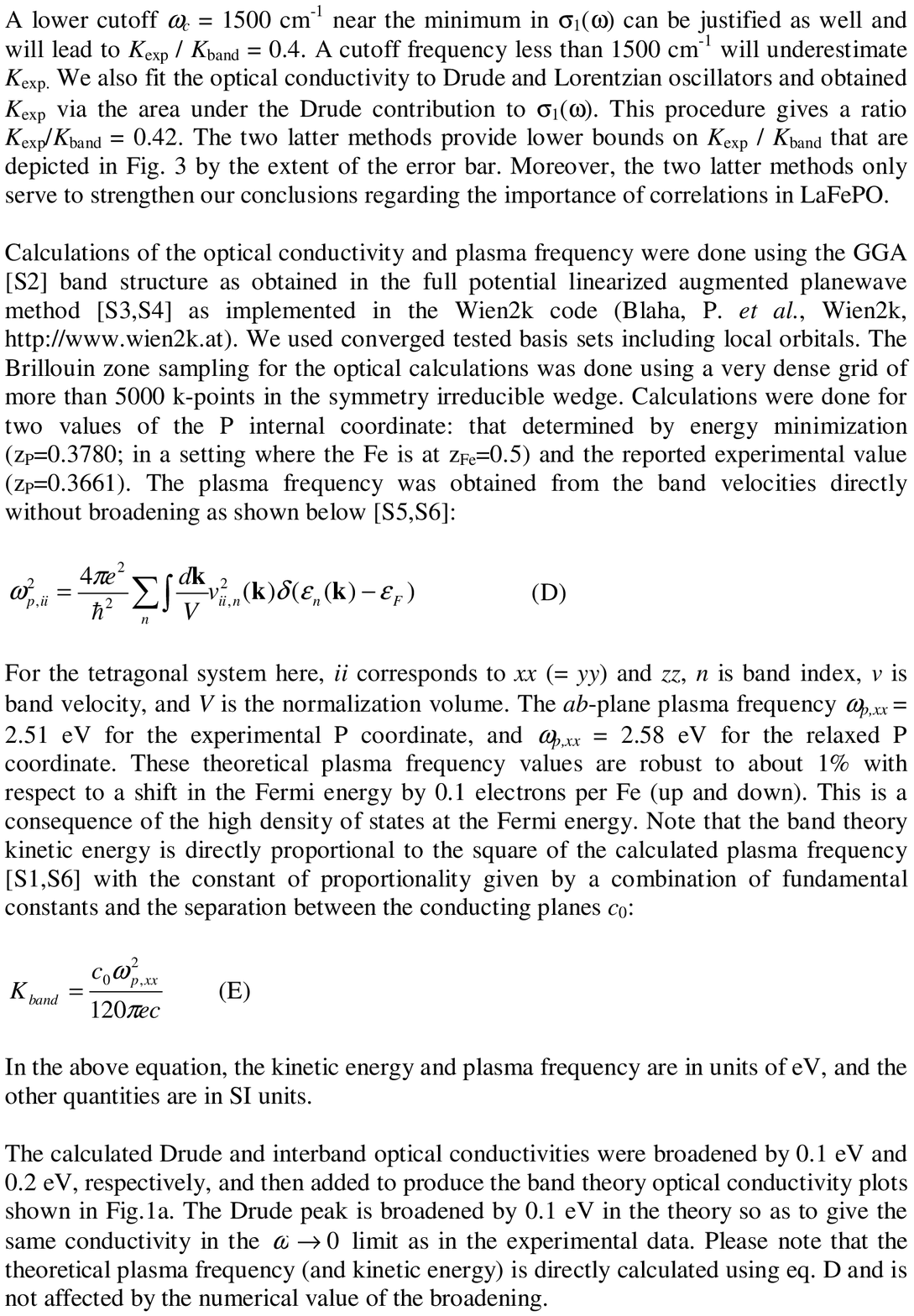}
\end{figure}

\newpage

\begin{figure}[t]
\epsfig{figure=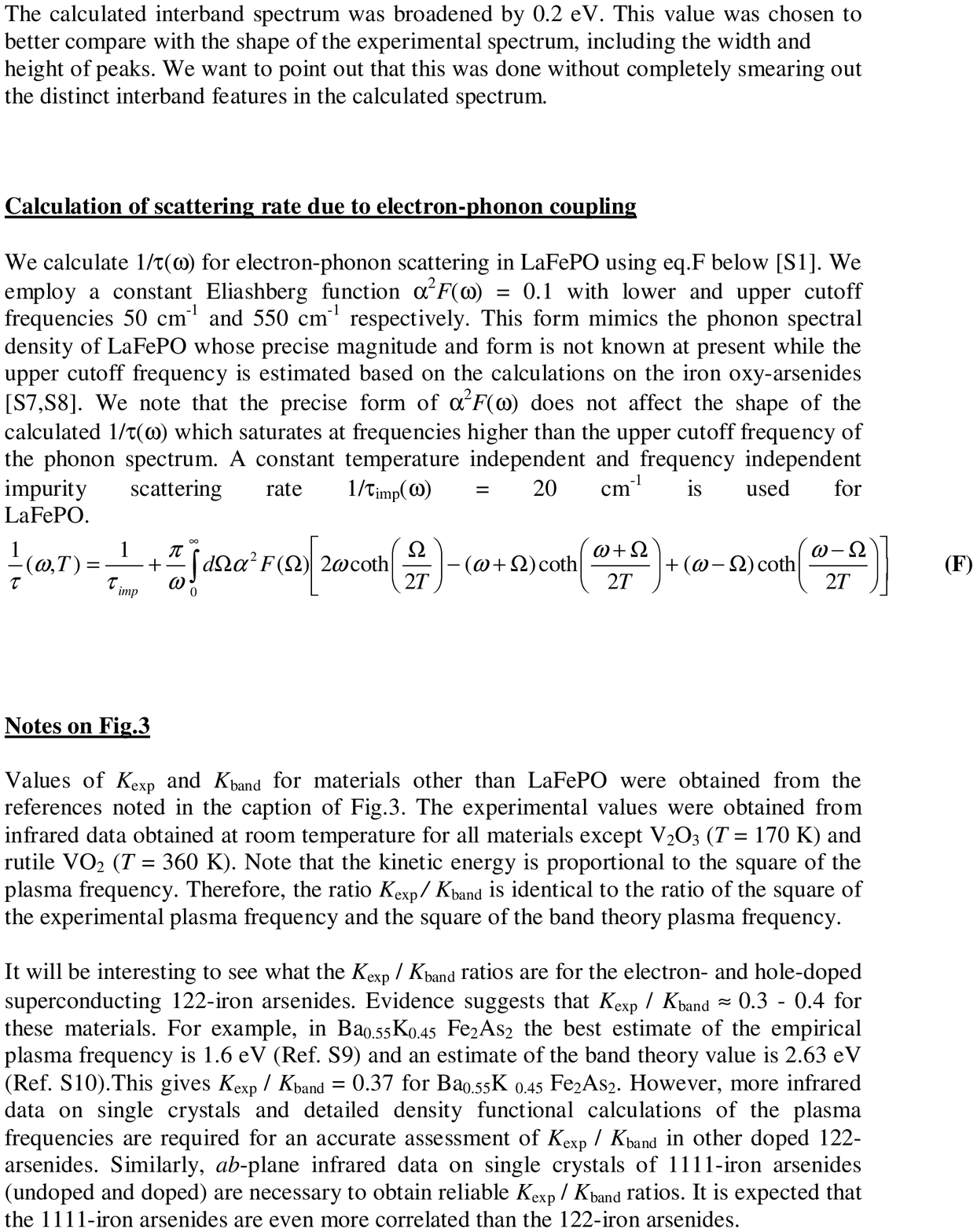}
\end{figure}

\newpage

\begin{figure}[t]
\epsfig{figure=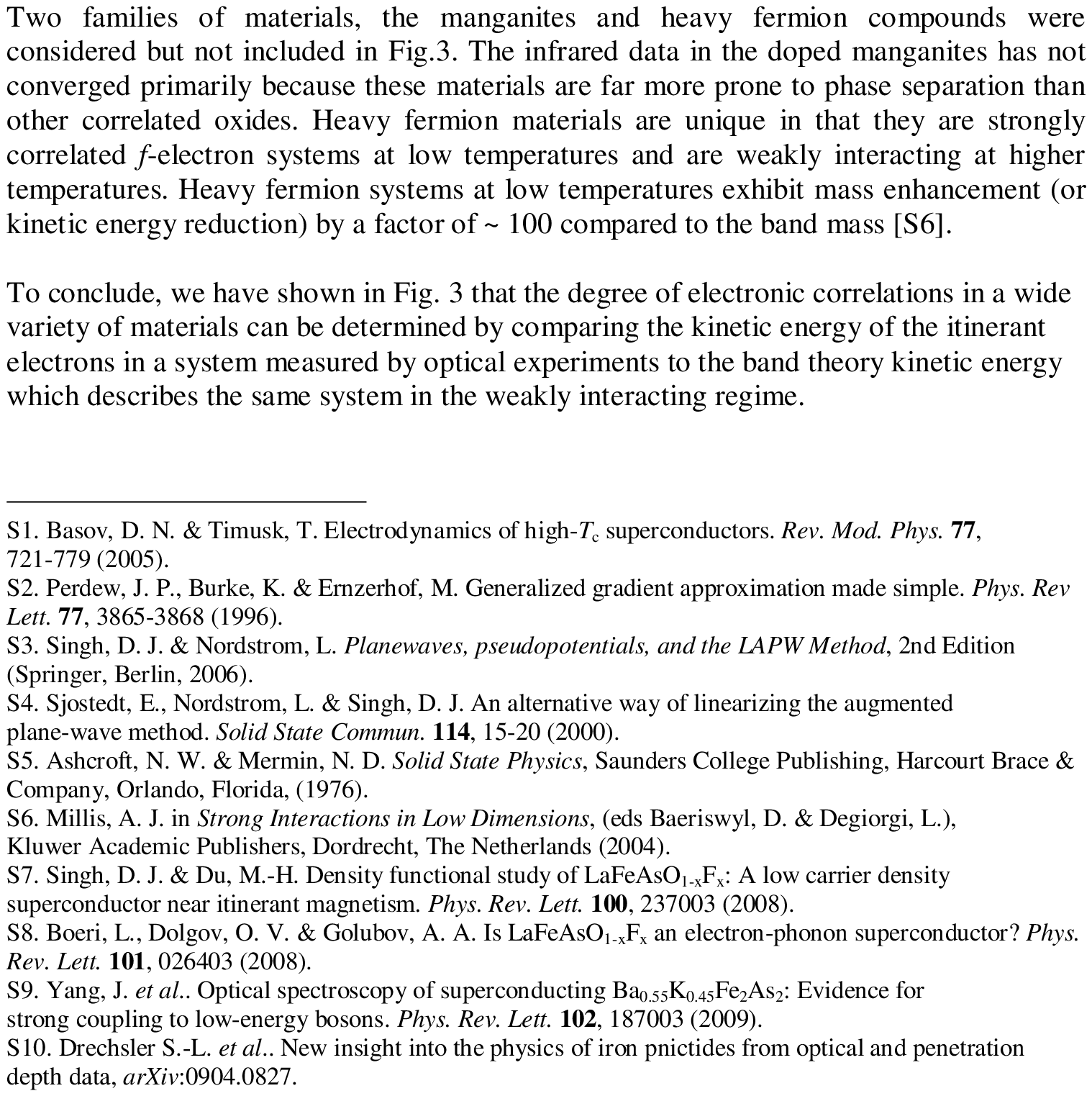}
\end{figure}

\end{document}